\documentclass[onecolumn]{revtex4}
\usepackage{graphicx,epsf}
\usepackage{color}
\usepackage{dcolumn}
\usepackage{amsfonts}
\usepackage{mathrsfs}
\usepackage{amsmath}

\usepackage{subfig}

\begin{document}

\title{Nonlinear excitation of fine-radial-scale zonal structures by toroidal Alfv\'en eigenmode}

\author{Zhiyong Qiu$^{1}$, Liu Chen$^{1, 2}$ and Fulvio Zonca$^{3, 1}$}

\affiliation{$^1$Institute for    Fusion Theory and Simulation and Department of Physics, Zhejiang University, Hangzhou, P.R.C\\
$^2$Department of   Physics and Astronomy,  University of California, Irvine CA 92697-4575, U.S.A.\\
$^3$ ENEA C. R. Frascati, C. P.
65-00044 Frascati, Italy}

\begin{abstract}
{
The set of   equations describing   nonlinear evolution of a single toroidal Alfv\'en eigenmode are derived, including both zero frequency zonal structure (ZFZS) generation and wave-particle phase space nonlinearities. The simplified case of neglecting wave-particle phase space nonlinearity is then investigated to focus on different roles of energetic particles and bulk plasmas on ZFZS generation. It is shown that energetic particles and bulk plasma play  dominant roles in ZFZS generation in different nonlinear stages, and the corresponding  processes are qualitatively different. Several properties of ZFZS generation, e.g., fine- vs. meso-scale, forced driven vs. spontaneous excitation, are clarified by the present analysis.
}
\end{abstract}

\maketitle

\section{Introduction}

Shear Alfv\'en waves (SAW) are expected to play an important role in future burning plasmas with  energetic particle (EP) population such as fusion-$\alpha$s significantly contributing to the overall plasma energy density \cite{LChenRMP2016}.  With  frequency comparable to the characteristic frequencies of EPs, and group velocities mainly along magnetic field lines, SAWs could be   driven unstable by   EPs \cite{YKolesnichenkoVAE1967,AMikhailovskiiSPJ1975,MRosenbluthPRL1975,LChenPoP1994} via resonant wave-particle interactions; leading to  EP transport and degradation of overall confinement, as reviewed in Ref. \cite{LChenRMP2016}.  As a consequence, understanding the nonlinear dynamics and saturation spectrum of SAWs is of crucial importance for the understanding of  future burning plasmas behavior.

There are two routes for the nonlinear saturation of Alfv\'en modes, i.e., nonlinear wave-particle and nonlinear wave-wave interactions \cite{LChenPoP2013}.
Wave-particle phase space nonlinearity \cite{FZoncaNJP2015}, e.g., wave-particle trapping,  describes the nonlinear distortion of the   EP distribution function; and leads to SAW saturation as the wave-particle trapping frequency, proportional to square root of the mode amplitude, is comparable with linear growth rate \cite{TOneilPoF1965,HBerkPoFB1990a,HBerkPoFB1990c,JZhuNF2014}.     On the other hand,  wave-wave coupling accounts for the transfer of   wave energy   from   unstable modes to the stable part of the fluctuation spectrum \cite{TSHahmPRL1995,LChenPRL2012}.
Thus, to correctly investigate the saturation of SAW, both wave-particle and wave-wave nonlinearities must be considered   on the same footing, with the intrinsic nonuniformity associated with toroidal geometry properly accounted for \cite{LChenPoP2013}.

Distinct by their different interactions with SAW continuum, there are two kinds of EP driven Alfv\'enic modes in toroidal devices, i.e., EP continuum modes (EPM)   and discrete Alfv\'en eigenmodes (AE). We note that, the general equations for studying EPM nonlinear physics \cite{LChenPoP1994}, including both zero frequency zonal structure (ZFZS) and phase space zonal structure (PSZS) generation,  are presented in Ref. \cite{FZoncaNJP2015}; though only PSZS are treated to give a clearer picture and discuss numerical simulations \cite{SBriguglioPoP2014}. Meanwhile, toroidal Alfv\'en eigenmode (TAE) \cite{CZChengAP1985,GFuPoFB1989}, excited inside the toroidicity-induced SAW continuum gap to minimize continuum damping, is one of most dangerous candidates for effectively scattering EPs.  In this work, we take TAE as an example, and derive the  set of   nonlinear equations for a single-$n$ TAE, and the $n=0$ ZFZS  \cite{MRosenbluthPRL1998,LChenPoP2000,LChenPRL2012} and  EP phase space nonlinearity. Spectral transfers by nonlinear scattering of the TAE spectrum are not considered here \cite{TSHahmPRL1995}. Thus, the present theoretical framework   provides a ``minimum" problem for self-consistent TAE studies.

Nonlinear excitation of  ZFZS  by SAWs is not favorable due to the properties of ``pure Alfv\'enic state"; that is, in uniform plasmas  under ideal MHD condition  the two dominant nonlinear terms, i.e., Maxwell (MX) and Reynolds stresses (RS), cancel each other so that a finite amplitude SAW, satisfying $\omega=\pm k_{\parallel}V_A$, can exist for a long time   without being affected by nonlinear processes.   By assumption, it is readily noted that   the pure Alfv\'enic state can be broken by, e.g., finite  compressibility \cite{ZQiuNF2016}, violation of ideal MHD constraint \cite{LChenEPL2011} and/or inhomogeneity/geometry \cite{LChenPRL2012,ZQiuEPL2013}. In the case of EP driven TAE considered here, we show that ZFZS generation can be due to the breaking of Alfv\'enic state by EPs \cite{ZQiuPoP2016} and/or toroidicity \cite{LChenPRL2012}.

ZFZS generation by TAE has been studied in several works. Numerical analyses of nonlinear dynamics of EP driven TAE are carried out by both hybrid code \cite{YTodoNF2010} and particle-in-cell code \cite{ZWang2016,ABiancalani2016} simulations, and found that zonal flow (ZF) is excited via essentially thresholdless forced driven process, with electrostatic ZF dominating over electromagnetic zonal current(ZC) and the ZF growth rate being twice the TAE  growth rate \cite{YTodoNF2010}.
On the other hand, Chen et al \cite{LChenPRL2012} investigated the nonlinear excitation of ZFZS by   TAE with a prescribed amplitude, and found that finite amplitude TAE can excite ZFZS via modulational instability at a rate proportional to the amplitude of the pump TAE. Zonal scalar potential is described by nonlinear vorticity equation  with the nonlinear drive from RS and MX unbalance, as  the pure Alfv\'enic state is broken by toroidicity. However, this process is limited by two facts. First of all, the net drive from unbalance of RS and MX is weak ($O(\epsilon)$) due to the small frequency mismatch of AE with respect to the SAW continuous spectrum; and, second, zonal scalar potential level is screened by enhanced neoclassical shielding due to magnetically trapped thermal plasma ions.
As a result, zonal current (zonal magnetic field) with lower excitation threshold could be preferentially excited in specific plasma equilibria, which, however, do not reflect typical experimental conditions of tokamak plasmas \cite{LChenPRL2012}.

It is shown in Ref. \cite{ZQiuPoP2016} that there is no conflict  between   analytical theory \cite{LChenPRL2012} and numerical simulations \cite{YTodoNF2010,ZWang2016}. The forced driven process \cite{YTodoNF2010,ZWang2016} is dominated by resonant EP contribution in the linear growth stage of the pump TAE; while the modulational instability \cite{LChenPRL2012} with a much slower time scale dominates when wave-particle interactions are weak. This  finding is novel, since it is usually believed that mode-mode coupling is dominated by bulk plasma non-resonant particles, while resonant particles play an important role in wave-particle nonlinearity.  In this paper, it will be demonstrated  that  the forced driven process \cite{ZQiuPoP2016,YTodoNF2010,ZWang2016} and the modulational instability \cite{LChenPRL2012}, which are respectively due to EP and thermal plasma nonlinear responses, dominate  the TAE nonlinear dynamics at different stages, respectively earlier and later.

The rest of the paper is organized as follows. In Sec. \ref{sec:model}, the theoretical model is presented, while  the set of nonlinear equations for the   nonlinear dynamics of TAE is derived in Sec. \ref{sec:nonlinear}. In Sec. \ref{sec:ZFZS_generation},  nonlinear generation of ZFZS is discussed in detail. Finally, a brief summary and discussions are  given in Sec. \ref{sec:summary}.

\section{Theoretical model}
\label{sec:model}

To derive the governing   equations for the nonlinear  evolution of the system, we take $\delta\phi$ and $\delta A_{\parallel}$ as the field variables. Here, $\delta\phi$ and $\delta A_{\parallel}$ are, respectively,  the scalar potential  and    component of the vector potential parallel to equilibrium magnetic field. An alternative field variable $\delta\psi\equiv \omega\delta A_{\parallel}/(ck_{\parallel})$ is also adopted for $n\neq0$ TAEs, and one has $\delta\psi=\delta\phi$ in the ideal MHD limit.  For the nonlinear interactions between TAE and ZFZS, we take $\delta\phi=\delta\phi_Z+\delta\phi_T$, with $\delta\phi_T=\delta\phi_0+\delta\phi_{0^*}$.  Here, $\delta\phi_0$ is the   TAE with positive real frequency, and $\delta\phi_{0^*}$ is its counterpart with negative real frequency. Note that, in  previous papers, TAEs are often separated into a constant amplitude pump and its upper/lower sidebands with much smaller amplitude to study the linear growth stage of the modulational instability. The more general approach adopted here  can be applied to recover earlier results  obtained  using pump and upper/lower sidebands as limiting case, as we will show in Sec. \ref{sec:MI_fine_scale}. The  well-known ballooning-mode decomposition in the $(r,\theta,\phi)$ field-aligned toroidal flux coordinates is assumed:
\begin{eqnarray}
\delta\phi_0=\hat{A}_0e^{i\int \hat{k}_{0,r} dr+i(n\phi-m_0\theta-\omega_0t)}\sum_j e^{-ij\theta}\Phi_0(x-j).
\end{eqnarray}
Here, $n$ is the toroidal mode number,  $(m=m_0+j)$ is the poloidal mode number with $m_0$ being its reference value satisfying $nq(r_0)=m_0$, $r_0$ is the plasma radial coordinate about which the TAE mode is assumed to be localized, $q(r)$ is the safety factor, $x=nq-m_0\simeq nq'(r_0)(r-r_0)$, $\Phi_0$ is the fine scale structure due to $k_{\parallel}$ radial dependence and magnetic shear,   $\hat{A}_0$ is the envelope amplitude and $\hat{k}_{0,r}\equiv nq'\theta_k$ is the radial envelope wavenumber in the ballooning  representation. For ZFZS   with the scalar potential dominated by $n=0, m=0$ components, and $n=0, m\simeq\pm1$ density perturbation \cite{LChenEPL2014}, we take
\begin{eqnarray}
\delta\phi_Z=\hat{A}_Ze^{i\int \hat{k}_Zdr-i\omega_Z t}\sum_m\Phi_Z\label{def:phiz}
\end{eqnarray}
with $\Phi_Z$ accounting for the fine radial structure \cite{ZQiuNF2016} due to nonlinear mode couplings, and $A_Z\equiv \hat{A}_Z\exp{(i\int \hat{k}_Zdr)}$ being the usual ``meso"-scale structure.
Note that  the summation of $m$ in the expression of $\delta\phi_Z$ indicates that the fine structure of $\delta\phi_Z$ \cite{ZQiuNF2016,ZQiuPoP2016} locates at the radial position of $\Phi_0(nq-m)$; i.e., $|nq-m|\simeq 1/2$ for TAE considered here. In fact, it is induced by the fine radial structures of Alfv\'en modes, which, in turn, is connected with their parallel mode structure because of the dependence of $k_{\parallel}$ on $r$.

The governing equations   can be derived from nonlinear gyrokinetic  vorticity equation \cite{LChenRMP2016}
\begin{eqnarray}
&&(c^2B)/(4\pi\omega^2)\partial_l(k^2_{\perp}/B)\partial_l \delta\psi_k+(e^2/T_i)\langle(1-J^2_k)F_0\rangle\delta\phi_k-\sum_s\left\langle(e_s/\omega)J_k\omega_d\delta H\right\rangle_k\nonumber\\
&=&-ic\Lambda_k\left[ c^2k''^2_{\perp}\partial_l\delta\psi_{k'}\partial_l\delta\psi_{k''}/(4\pi\omega_{k'}\omega_{k''})+\left\langle e(J_kJ_{k'}-J_{k''})\delta L_{k'}\delta H_{k''}\right\rangle\right]/(\omega_kB_0),
\label{eq:vorticityequation}
\end{eqnarray}
where the two explicitly nonlinear terms on the right hand side are, respectively,  MX and RS,  $\langle\cdots\rangle$ indicates velocity space integration,  the subscripts $s=i, e, E$ denote  particle species (thermal ions, electrons and EPs), and $\Lambda_k\equiv \sum_{\mathbf{k}'+\mathbf{k}''=\mathbf{k}}\hat{\mathbf{b}}\cdot\mathbf{k}''\times\mathbf{k}'$.
Here, $\mathbf{k}$ are defined as the operators  for spatial derivatives, and we have
$\mathbf{k} \delta\phi \equiv[k_{\parallel}\mathbf{b}+ k_{\theta}\hat{\mathbf{\theta}}+\left(\hat{k}_r-inq'\partial_x \ln\Phi\right)\hat{\mathbf{r}}]\delta\phi$. Thus, $nq'\partial_x\ln\Phi$ is related to the radial derivative of the fine radial structure, while $\hat{k_r}$ is the radial envelope wave number accounting for the typical ``meso"-scale envelope structures.
The nonadiabatic particle response is derived from the nonlinear gyrokinetic equation \cite{EFriemanPoF1982}:
\begin{eqnarray}
\left(-i\omega+v_{\parallel}\partial_l+i\omega_d\right)\delta H_k&=&-i(e_s/m)QF_0J_k\delta L_k-(c/B_0)\Lambda_kJ_{k'}\delta L_{k'}\delta H_{k''}\label{eq:NLGKE}.
\end{eqnarray}
Here, $QF_0=(\omega\partial_E-\omega_*)F_0$ with $E=v^2/2$, $\omega_*F_0=\mathbf{k}\cdot\mathbf{b}\times\nabla F_0/\Omega$.  Furthermore,  $\omega_d=(v^2_{\perp}+2
v^2_{\parallel})/(2 \Omega R_0)\left(k_r\sin\theta+k_{\theta}\cos\theta\right)$   for a circular cross section large aspect ratio tokamak,  $l$ is the length along the equilibrium magnetic field line, $J_k=J_0(k_{\perp}\rho_L)$ with $J_0$ being the Bessel function accounting for finite Larmor radius effects,  $\rho_L\equiv mcv_{\perp}/(eB)$ is the Larmor radius,   $\delta L=\delta\phi-v_{\parallel}\delta A_{\parallel}/c$; and other notations are standard.

To close the system, we need another equation. Here, we take parallel component of   Ohm's law:
\begin{eqnarray}
\delta E_{\parallel,k} +\sum_{\mathbf{k}'+\mathbf{k}''=\mathbf{k}}\hat{\mathbf{b}}\cdot\delta \mathbf{u}_{k'}\times\delta\mathbf{B}_{k''}/c=0\label{eq:nl_Ohm}.
\end{eqnarray}
Here, $\delta \mathbf{u}$ is the $\mathbf{E}\times\mathbf{B}$ drift velocity. Equation (\ref{eq:nl_Ohm}) is equivalent to the usual quasi-neutrality condition, neglecting $O(k^2_{\perp}\rho^2_i)$ terms.  The nonlinear equations describing the self-consistent evolution of TAE can then be derived from equations (\ref{eq:vorticityequation}), (\ref{eq:NLGKE}) and (\ref{eq:nl_Ohm}).

\section{Nonlinear TAE saturation by ZFZS and PSZS}
\label{sec:nonlinear}

In this section, based on the     general equations  (\ref{eq:vorticityequation}) - (\ref{eq:nl_Ohm}),  we will derive the   nonlinear equations describing the self-consistent evolution of a single toroidal mode number TAE, including both $n=0$ ZFZS and PSZS generation, and the feedback of ZFZS and PSZS on TAE.  For simplicity of discussion, well circulating EPs are assumed. However, the theoretical framework presented here is general, and can be applied to both circulating and trapped EPs. Extension to different Alfv\'en modes, e.g., beta-induced AE (BAE) \cite{WHeidbrinkPRL1993,FZoncaPPCF1996} or EPM \cite{LChenPoP1994}, is straightforward.

\subsection{Nonlinear ZFZS equations}
\label{subsec:ZFZS}

Due to their different drift orbit/Larmor radius sizes, EPs, thermal ions and electrons contribute to different    terms in the vorticity equation, where $\delta\phi_Z$ is determined. Thus, our derivations can be greatly simplified with the guidance of the multiple scale properties of TAE mode structure \cite{FZoncaPPCF2015}. For the sake of clarity,  MX, RS and the curvature coupling term (CCT)   are derived one by one in the following.

MX is dominated by current-carrying electrons. From equation (\ref{eq:vorticityequation}) applied to ZFZS, one has
\begin{eqnarray}
\mbox{MX}&=&-i\frac{c}{B_0}\frac{1}{\omega_Z}\overline{\sum_{\mathbf{k}}\hat{\mathbf{b}}\cdot\mathbf{k}''\times\mathbf{k}'\frac{c^2k''^2_{\perp}\partial_l\delta\psi_{k'}\partial_l\delta\psi_{k''}}{4\pi\omega'\omega''}}\nonumber\\
&=&-i\frac{c}{B_0}\frac{1}{\omega_Z}\frac{c^2}{4\pi}\frac{k^2_{\parallel,0}}{\omega^2_0}\hat{\mathbf{b}}\cdot\mathbf{k}_0\times\mathbf{k}_{0^*}\left(k^2_{\perp,0}-k^2_{\perp,0^*}\right)\overline{\delta\psi_0\delta\psi_{0^*}}\nonumber\\
&=&-i\frac{c}{B_0}\frac{1}{\omega_Z}\frac{c^2}{4\pi}\frac{k^2_{\parallel,0}}{\omega^2_0}k_{\theta,0}(k_{r,0}+k_{r,0^*})^2 (k_{r,0^*}-k_{r,0})\overline{\delta\psi_0\delta\psi_{0^*}}.\label{eq:MX1}
\end{eqnarray}
Here, $\overline{(\cdots)}$ denotes surface averaging, and equation ($\ref{eq:MX1}$) is derived noting $|k_{\parallel,0}|\simeq 1/(2qR_0)$ for TAEs. Noting that $k_r$ is the operator for radial derivative, we then have
\begin{eqnarray}
\mbox{MX}&=&-\frac{c}{B_0}\frac{1}{\omega_Z}\frac{c^2}{4\pi}\frac{k^2_{\parallel,0}}{\omega^2_0}k_{\theta,0}\frac{\partial^2}{\partial r^2}\left(\hat{F}|\hat{A}_0|^2\sum_m|\Phi_0|^2\right).\label{eq:MX}
\end{eqnarray}
Here, $\hat{F}\equiv i(\hat{k}_{r,0}-\hat{k}_{r,0^*})+\partial_r\ln\Phi_0-\partial_r\ln\Phi_{0^*}$, with $\hat{k}_{r,0}-\hat{k}_{r,0^*}$ accounting for radial envelope modulation \cite{LChenPRL2012} and $\partial_r\ln\Phi_0-\partial_r\ln\Phi_{0^*}$ related with fine radial structures of TAE  \cite{ZQiuNF2016}. The occurrence of $\partial^2_r$ in the MX expression demonstrates that this nonlinearity becomes most important at radial locations  where fine TAE radial structures are predominant; that is in the inertial layer of AEs.

Reynolds stress nonlinearity is also most important in the inertial layer, since it can be noted that RS contributes only when $k_{\perp}\rho_L\lesssim 1$, which is dominated by thermal ions response, while EP ($k_{\perp}\rho_L\gg1$) and thermal electron  ($k_{\perp}\rho_L\ll1$) contribution to RS is negligible.
With the $|\omega_0|\gg|k_{\parallel}v_{\parallel}|, |\omega_d|$   ordering, leading order thermal ion response to TAE can be derived as $\delta H^L_{T,i}\simeq(e/T_i)F_0J_k\delta\phi_T$. Substituting into vorticity equation for the ZFZS, we then have \cite{FZoncaPoP2004,ZQiuPoP2014}
\begin{eqnarray}
\mbox{RS}&=&-i\frac{c}{B_0}\frac{1}{\omega_Z}\overline{\sum_{\mathbf{k}}\hat{\mathbf{b}}\cdot\mathbf{k}''\times\mathbf{k}'\left\langle e\left(J_kJ_{k'}-J_{k''}\right)\delta L_{k'}\delta H_{k'',i}\right\rangle}\nonumber\\
&=&-i\frac{c}{B_0}\frac{e}{\omega_Z}\hat{\mathbf{b}}\cdot\mathbf{k}''\times\mathbf{k}'\overline{\left\langle (J_k-1)\left(J_{k'}\delta L_{k'}\delta H_{k'',i}-J_{k''}\delta L_{k''}\delta H_{k',i}\right)\right.}\nonumber\\
&&\hspace*{8em}\overline{\left.+\left(J_{k'}-J_{k''}\right)\left(\delta L_{k'}\delta H_{k'',i}+\delta L_{k''}\delta H_{k',i}\right)\right\rangle}.
\end{eqnarray}
For thermal ions with $|k_{\perp}\rho_i|\lesssim1$ in the inertial layer, the second term on the right hand side of RS dominates. Therefore,
\begin{eqnarray}
\mbox{RS}&=&\frac{c}{B_0}\frac{1}{\omega_Z}\frac{n_0e^2}{T_i}k_{\theta,0}\rho^2_i\frac{\partial^2}{\partial r^2} \left(\hat{F}|\hat{A}_0|^2\sum_m|\Phi_0|^2\right). \label{eq:RS}
\end{eqnarray}

Combining equations (\ref{eq:MX}) and (\ref{eq:RS}), We then have
\begin{eqnarray}
\mbox{RS+ MX}&=&-\frac{1}{2}\frac{c}{B_0}\frac{n_0e^2}{T_i}k_{\theta}\rho^2_i\frac{1}{\omega_Z}\left(1-\frac{k^2_{\parallel,0}v^2_A}{\omega^2}\right)\frac{\partial^2}{\partial r^2}\left(\hat{F}|\hat{A}_0|^2\sum_m|\Phi_0|^2\right). \label{eq:RS_and_MX}
\end{eqnarray}
Note that equation (\ref{eq:RS_and_MX}) reproduces equation (5) of Ref. \citenum{LChenPRL2012} if one neglects the resonant particle effects ($\partial_r\ln\Phi_0-c.c.$ in $\hat{F}$), and separates  TAE sidebands due to radial envelope  modulation of ZFZS from the pump TAE.
The finite coupling comes from toroidicity ($1-k^2_{\parallel}V^2_A/\omega^2_0 \neq0$, breaking of Alfv\'enic state) and $\hat{F}\neq0$ due to either envelope modulation \cite{LChenPRL2012} or wave-particle resonances \cite{ZQiuPoP2016}.

Energetic particles,   with $|k_{\perp}\rho_{d,E}|\gg1$ in the inertial layer \cite{FZoncaPPCF2015}, do not contribute to RS  \cite{LChenPRL2012}.   Here,  $\rho_{d,E}$ is the magnetic drift orbit width. EP nonlinearity enters implicitly   in the ideal region via the CCT contribution due to the  nonlinear EP contribution to ZFZS. Another reason for EP contribution to be favored in the CCT is that  CCT is related to the  particles pressure instead of density.
Noting $|\omega_{*,E}|\gg|\omega_0|$ for typical EP driven TAEs \cite{GFuPoFB1989,HBiglariPoFB1992}, the  nonlinear gyrokinetic equation for EP response to TAE can be written as \cite{FZoncaNJP2015}
\begin{eqnarray}
\left(-i\omega+v_{\parallel}\partial_l+i\omega_d-ck_Zk_{\theta}J_Z\delta L_Z/B_0\right)\delta H_0&=&-i(c/B_0)J_0\delta L_0k_{\theta}\partial_r \bar{F}_0.\label{eq:NLGKE_renormalized}
\end{eqnarray}
Here, $k_Z$ stands for $k_r$ of the ZFZS response, $J_Z=J_0(k_Z\rho_L)$ and $\delta L_Z=\delta\phi_Z-(v_{\parallel}/c)\delta A_{\parallel,Z}$. Thus,  $ck_Zk_{\theta}J_Z\delta L_Z/B_0$ corresponds to the scattering of EP orbit by slowly varying ZS. Meanwhile, $\bar{F}_0=F_0+\overline{\delta H^{NL}_Z}$ is the  ``time evolving"  equilibrium EP distribution function, and its expression will be derived in Sec. \ref{subsec:PSZS}. Note that,   compared to equation (\ref{eq:NLGKE}), the current equation includes only the surface averaged component of $\delta H^{NL}_Z$, since $\overline{\delta H^{NL}_Z}$ dominates    the  principal  series of secular terms in the perturbation expansion \cite{FZoncaNJP2015}.
$\overline{\delta H^{NL}_Z}$ is the   phase space zonal structure, and reproduces hole-clump pair creation in the adiabatic limit \cite{HBerkPoFB1990a,HBerkPoFB1990b,HBerkPoFB1990c}. The free energy in velocity space (the first term in $QF_0$, defined below equation (\ref{eq:NLGKE})) is neglected with respect to that in configuration space (the second term in $QF_0$), due to the $|\omega_{*,E}|\gg|\omega_0|$ ordering. Equation (\ref{eq:NLGKE_renormalized}) can be derived from equation (15) of Ref. \cite{FZoncaNJP2015}, assuming $|\omega_{*,E}|\gg|\omega_0|$ ordering and circular cross section.  EP response to TAE can be derived as \cite{HBiglariPoFB1992,ZQiuPoP2016}
\begin{eqnarray}
\delta H^F_0=\frac{c}{B_0}k_{\theta}J_0\delta L_0\partial_r\bar{F}_0e^{i\lambda_d}\sum_l\frac{J^2_l(\hat{\lambda}_d)e^{il(\theta-\theta_0)}}{\omega_0-k_{\parallel,0}v_{\parallel}-l\omega_{tr}-ick_Zk_{\theta}J_Z\delta L_Z/B_0}. \label{eq:nl_EP_TAE}
\end{eqnarray}
Here, $\lambda_d=k_Z\rho_d$ with $\partial_{\theta}\rho_d=-\omega_d/(k_Z\omega_{tr})$.  This expression has a form   similar to the linear  EP response to TAE derived in Ref. \cite{ZQiuPoP2016}, but with a nonlinear propagator and a modified $\bar{F}_0$.
This is the fully nonlinear EP response to a single toroidal mode number TAE, including the nonlinear scattering of EP orbit   by ZS and EP ``equilibrium" distribution function modification by TAE (transport). It also bears the information of TAE frequency sweeping due to EP energy or $P_{\phi}$ variations  by TAE and phase locking between TAE and resonant EPs \cite{LChenRMP2016,FZoncaNJP2015}.   We note that, the $ick_Zk_{\theta}J_Z\delta L_Z/B_0$ term in the nonlinear propagator  may contribute to resonance detuning or resonance broadening, depending on the mechanism of ZS generation.
The nonlinear TAE equation can then be   derived by substituting equation (\ref{eq:nl_EP_TAE}) into vorticity equation, and this will be done in Sec. \ref{subsec:NL_TAE}.

Nonlinear EP contribution to ZFZS can be derived by transforming into the drift orbit center coordinates \cite{FZoncaNF2005}. Taking $\delta H^{NL}_Z=e^{i\lambda_{dZ}}\delta H^{NL}_{dZ}$, we then have
\begin{eqnarray}
\left(\partial_t+\omega_{tr}\partial_{\theta}\right)\delta H^{NL}_{dZ}=-\frac{c}{B}e^{-i\lambda_{dZ}}\Lambda_Z J_0(\gamma_{k'})\delta L_{k'}\delta H_{k''}.
\end{eqnarray}
Separating $\delta H^{NL}_{dZ}=\overline{\delta H^{NL}_{dZ}}+\widetilde{\delta H^{NL}_{dZ}}$, with $\widetilde{(\cdots)}$ denoting   poloidally varying component, and noting that $|\widetilde{\delta H^{NL}_{dZ,E}}/\overline{\delta H^{NL}_{dZ,E}}|\sim |\omega_Z/\omega_{tr,E}|\ll1$, we then have
\begin{eqnarray}
\partial_t\overline{\delta H^{NL}_{dZ}}&=&-\frac{c}{B_0}\overline{e^{-i\lambda_{dZ}}\Lambda_Z J_0(\gamma_{k'})\delta L_{k'}\delta H_{k''}},\label{NLdc}\\
\omega_{tr}\partial_{\theta}\widetilde{\delta H^{NL}_{dZ}}&=&-\frac{c}{B_0}\left[e^{-i\lambda_{dZ}}\Lambda_Z J_0(\gamma_{k'})\delta L_{k'}\delta H_{k''}\right]_{AC}\label{NLac}.
\end{eqnarray}
Here, the subscript ``AC" denotes $m\neq0$ component, and $(\cdots)_{AC}=\widetilde{(\cdots)}$.
Although  $|\widetilde{\delta H^{NL}_{dZ,E}}/\overline{\delta H^{NL}_{dZ,E}}|\sim |\omega_Z/\omega_{tr,E}|\ll1$, the dominant EP contribution to ZFZS generation  comes from $\widetilde{\delta H^{NL}_{dZ,E}}$ since it enters vorticity equation via coupling with geodesic curvature. On the other hand, $\overline{\delta H^{NL}_{dZ,E}}$, which is the phase space zonal structure response, dominates the ZFZS feedback onto the AE fluctuation spectrum. That is, it will dominate the nonlinear wave-particle interaction \cite{LChenRMP2016,FZoncaNJP2015}, as we discussed below equation (\ref{eq:NLGKE_renormalized}) and we  will further discuss it   in Sec. \ref{subsec:PSZS}.
Noting that $\omega_{dZ}=\omega_{tr}\partial_{\theta}\lambda_{dZ}$,   the EP contribution to ZFZS generation via CCT, after integration by parts, can be rewritten as
\begin{eqnarray}
\mbox{CCT}&=&\left\langle\overline{\frac{e}{\omega}J_Z\omega_d\delta H^{NL}_Z}\right\rangle\nonumber\\
&=&-\frac{i}{2\pi}\frac{e}{\omega}\left\langle J_Z\int d\theta e^{i\lambda_{dZ}}\omega_{tr}\partial_{\theta}\widetilde{\delta H^{NL}_{dZ}}\right\rangle.\label{CCT1}
\end{eqnarray}
Substituting equation (\ref{NLac}) into equation (\ref{CCT1}), and noting  that $\overline{A \widetilde{B}}=\overline{\widetilde{A}B}$ and $\widetilde{e^{i\lambda_{dZ}}}=e^{i\lambda_{dZ}}-J_0(\hat{\lambda}_{dZ})$, we then have
\begin{eqnarray}
\mbox{CCT}&=&\frac{i}{2\pi}\frac{c}{B_0}\frac{e}{\omega}\left\langle J_Z\left[\underbrace{\int d\theta\Lambda_ZJ_{k'}\delta L_{k'}\delta H_{k''}}_{\mathscr{A}}\right.\right.\nonumber\\
&-&\left.\left.J_0(\hat{\lambda}_{dZ})\underbrace{\int d\theta e^{-i\lambda_{dZ}}\Lambda_ZJ_{k'}\delta L_{k'}\delta H_{k''}}_{\mathscr{B}}\right]\right\rangle.\label{eq:AB}
\end{eqnarray}
The $\mathscr{A}$ and $\mathscr{B}$ terms will be treated separately and rewritten more explicitly in the following, adopting a weak field perturbation expansion.

Using linearized   EP responses in the nonlinear terms (i.e., the linear expression for $\delta H_{k''}$ in the expression above), and ignoring the weak non-local coupling between two poloidal harmonics located at different radial positions, we then have
\begin{eqnarray}
\mathscr{A}&=&-\hat{H}\int d\theta\left[J_{0^*}\delta L_{0^*}\delta H_0-J_0\delta L_0\delta H_{0^*}\right]\nonumber\\
&=&2\pi \hat{H}\frac{e}{m}J_0 J_{0^*}\hat{A}_0\hat{A}_{0^*}\nonumber\\
&&\times\sum_m|\Phi_0|^2
\left(1-\frac{k_{\parallel}v_{\parallel}}{\omega}\right)_0\left(1-\frac{k_{\parallel}v_{\parallel}}{\omega}\right)_{0^*} Q_0F_0 \nonumber\\
&&\times\sum_l\left[\frac{J^2_l(\hat{\lambda}_{d0})}{\omega_0-k_{\parallel,0}v_{\parallel}-l\omega_{tr}-ick_Zk_{\theta,0}J_Z\delta L_Z/B_0}\right.\nonumber\\
&&+\left.\frac{J^2_l(\hat{\lambda}_{d0^*})}{\omega_{0^*}-k_{\parallel,0^*}v_{\parallel}-l\omega_{tr}-ick_Zk_{\theta,0^*}J_Z\delta L_Z/B_0}\right].\nonumber\\
\label{eq:A1}
\end{eqnarray}
Here, $\hat{H}=k_{\theta}(k_{r,0}+k_{r,0^*})$. In deriving equation (\ref{eq:A1}), the ideal MHD condition for TAE ($\delta\phi^L\simeq\delta\psi^L$)  is applied to simplify $\delta L_0$ and $\delta L_{0^*}$ \cite{ZQiuEPL2013}.

Assuming that dominant contribution comes from resonant EPs,  we then have
\begin{eqnarray}
\mathscr{A}&=&-2i\pi^2\hat{H}\frac{e}{m}J_0J_{0^*}Q_0F_0\frac{1}{\omega^2_0}\nonumber\\
&&\times\sum_l l^2\omega^2_{tr}\left(J^2_l(\hat{\lambda}_{d0})+J^2_l(\hat{\lambda}_{d0^*})\right)\nonumber\\
&&\times\delta (\omega_0-k_{\parallel,0}v_{\parallel}-l\omega_{tr}-ick_Zk_{\theta,0}J_Z\delta L_Z/B_0)\nonumber\\
&&\times|\hat{A}_0|^2\sum_m|\Phi_0|^2.\label{eq:A}
\end{eqnarray}
In deriving equation (\ref{eq:A}), the resonance condition is applied to simplify $\delta L_k$ (i.e., $\omega-k_{\parallel,0}v_{\parallel}\simeq l\omega_{tr}$). Equation (\ref{eq:A}) suggests that, in general, the  effects of ZS scattering can be  resonance detuning (nonlinear frequency shift) and/or resonance broadening.   For example, as will be shown later, when $\delta L_Z$ is imaginary and $k_Z$ predominantly real, as in the linear growth stage where ZS is forced driven by resonant EP effects \cite{ZQiuPoP2016}, ZS scattering on wave particle resonant interaction with TAE predominantly enhances resonance detuning. Similarly, as will be shown later, while the spontaneously excited ZF/ZC structures are real, and the effect on TAE resonances with EPs is predominantly resonance broadening. However, as we clarify in Sec. \ref{sec:ZFZS_generation} below, the CCT term due to EP contribution can be ignored with respect to MX and RS,  when ZF/ZC are spontaneously excited. Thus, equation (\ref{eq:A}) suggests that EP contribute predominantly to ZFZS forced driven excitation  and, thus, they enhance resonance detuning. Note that the CCT is due to  resonant wave-particle  interactions and  the finite orbit width  (FOW) effects via $l\neq0$ transit harmonics.   One would then expect, compared to the well-circulating EPs assumed here, that trapped EPs may further enhance the nonlinear couplings    due to their  relatively large bounce orbits.

The $\mathscr{B}$ term can be manipulated and rewritten   similarly. Substituting   EP response ($\delta H_{k''}$) into $\mathscr{B}$, and noting $k_Z=k_{r,0}+k_{r,0^*}$,  we obtain \cite{ZQiuPoP2016}:
\begin{eqnarray}
\mathscr{B}&=&\hat{H}\frac{e}{m}J_0J_{0^*}\int d\theta e^{-i\lambda_{dZ}}\delta L_0\delta L_{0^*}Q_0F_0\nonumber\\
&&\times\left[e^{-i\lambda_{d0}}\sum_l\frac{J_l(k_{\perp,0}\hat{\rho}_d)e^{il(\theta-\theta_0)}}{\omega_0-k_{\parallel,0}v_{\parallel}-l\omega_{tr} -ick_Zk_{\theta,0}J_Z\delta L_Z/B_0}\right.\nonumber\\
&&\hspace*{1.5em}\left.+e^{i\lambda_{d0^*}}\sum_l\frac{(-1)^lJ_l(k_{\perp,0^*}\hat{\rho}_d)e^{il(\theta+\theta_{0^*})}}{\omega_{0^*} -k_{\parallel,0^*}v_{\parallel}-l\omega_{tr}-ick_Zk_{\theta,0^*}J_Z\delta L_Z/B_0}\right]\nonumber\\
&=&0.\label{eq:B}
\end{eqnarray}

Collecting results from equations (\ref{eq:A}) and (\ref{eq:B}), assuming $|k_{\perp}\rho_{d,E}|\ll1$, and keeping only $l=\pm1$ transit resonances, equation (\ref{eq:AB}) finally becomes
\begin{eqnarray}
\mbox{CCT}&=&-\frac{\pi}{4}\hat{H}\frac{c}{B_0}\frac{e^2}{m}\frac{n_{0E}}{\omega_Z}\frac{k^2_{\perp}}{\omega^2_0}\hat{G}|\hat{A}_0|^2\sum_m|\Phi_0|^2.\nonumber
\end{eqnarray}
Here, $\hat{G}$ comes from resonant EP, and is defined as
\begin{eqnarray}
\hat{G}&\equiv&\left\langle \omega_{*,E}\hat{v}^2_{d,E}(\bar{F}_{0E}/n_{0E})\right.\nonumber\\
&&\left.\times\left(\delta(\omega_0-k_{\parallel}v_{\parallel}-\omega_{tr}-ick_Zk_{\theta,0}J_Z\delta L_Z/B_0)+\delta(\omega_0-k_{\parallel}v_{\parallel}+\omega_{tr}-ick_Zk_{\theta,0}J_Z\delta L_Z/B_0)\right)\right\rangle\nonumber.
\end{eqnarray}
Ignoring the term related to ZS induced scattering $ick_Zk_{\theta}J_Z\delta L_Z/B_0$,  $\hat{G}$ is proportional to the resonant EP contribution. Thus, the CCT term is proportional to the effective ``linear"  growth rate of TAE, and is important only in the linear growth stage of TAE.   In the expression of $\hat{G}$, the FLR effects are consistently ignored due to the $k_{\perp}\rho_{d,E}\ll1$ assumption.

The nonlinear vorticity equation for ZFZS can then be derived as
\begin{eqnarray}
\noindent\omega_Z\hat{\chi}_{iZ}\delta\phi_Z=-\frac{\pi}{4}\frac{c}{B_0}\left(\frac{k_{\theta}k^2_{\perp}}{k_Z}\frac{1}{\omega^2_0}\frac{n_{0E}}{n_0}
\frac{T_i}{m_i\rho^2_i}\hat{G} -\frac{2}{\pi}k_{\theta}\left(1-\frac{k^2_{\parallel}V^2_A}{\omega^2_0}\right)\hat{F}\right)|\hat{A}_0|^2\sum_m|\Phi_0|^2,\label{eq:ZFVorticity}
\end{eqnarray}
with the first term on the right hand side originating from resonant EP contribution to CCT in the ideal region, and the second term from RS\&MX of thermal plasma contribution in inertial layer.
The CCT by EPs is much larger than RS\&MX by  $O(n_{E,R}\hat{\omega}_{*E}q^2/(n_0\omega_0\epsilon))$, and dominates in the linear growth stage of TAE. Here, $n_{E,R}$ is the ``number" of resonant EPs. On the other hand, as $n_{E,R}$ decreases due to, e.g., wave-particle phase space nonlinearity,   RS\&MX may take over the long time scale nonlinear behavior. Thus, the physics investigated in Ref. \cite{LChenPRL2012} and Refs. \cite{YTodoNF2010,ZWang2016} occur at different stages of the nonlinear  dynamics. We will discuss the differences between these two processes in more detail in Sec. \ref{sec:ZFZS_generation}.

The equation for zonal magnetic field  can be derived from the parallel component of nonlinear Ohm's law \cite{LChenPRL2012}.
Noting $\delta E_{\parallel,Z}=-\partial_l\delta\phi_Z-\partial_t\delta A_{\parallel,Z}/c$, $\delta\mathbf{B}=\nabla\times\delta A_{\parallel}\mathbf{b}$ and $\delta \mathbf{u}_{\perp}=c\nabla\delta\phi\times\mathbf{b}/B$, the zonal component of equation (\ref{eq:nl_Ohm}) is then
\begin{eqnarray}
\partial_t\delta A_{\parallel,Z}&=&-\frac{c}{B}\mathbf{b}\cdot\nabla_{\perp}\delta A_{\parallel}\times \nabla_{\perp}\delta\phi\nonumber\\
&=&i\frac{c}{B_0}k_{\theta,0}\frac{\partial}{\partial r}\left(\delta A_{\parallel,0}\delta\phi_{0^*}-\delta\phi_0\delta A_{\parallel,0^*}\right).\nonumber
\end{eqnarray}

Taking $\delta\phi=\delta\psi$ for TAEs in the inertial layer, and noting that $|\partial_r \ln k_{\parallel}|\ll |k_r|$ for TAEs, we then have
\begin{eqnarray}
\partial_t\delta A_{\parallel,Z}=i\frac{c}{B_0}k_{\theta,0}k_{\parallel_0}\left(\frac{1}{\omega_0}+\frac{1}{\omega_{0^*}}\right)\frac{\partial}{\partial r} |\delta\phi|^2.\nonumber
\end{eqnarray}
Noting that $\omega_0=\omega_{0r}+i\partial_t$, we obtain
\begin{eqnarray}
\delta\psi_Z=-\frac{1}{B_0}\frac{k_{\theta,0}}{\omega_{0}}\frac{\partial}{\partial r}\left(|\hat{A}_0|^2\sum_m|\Phi_0|^2\right).   \label{eq:zonalcurrent}
\end{eqnarray}
Here,  $\delta\psi_Z\equiv (\omega_0/ck_{\parallel,0})\delta A_{\parallel,Z}$.

\subsection{Nonlinear TAE equations}
\label{subsec:NL_TAE}

In the vorticity equation of TAE, the nonlinear terms contains the CCT dominated by EP response, and RS\&MX responses due to bulk plasmas discussed in Ref. \cite{LChenPRL2012}. Substituting equation (\ref{eq:nl_EP_TAE}) into (\ref{eq:vorticityequation}), one then obtains
\begin{eqnarray}
&&-k^2_{\parallel}\delta\psi_0+\frac{\omega^2_0}{V^2_A}\delta\phi_0-\frac{\omega^2_G}{V^2_A}\delta\phi_0-\frac{4\pi\omega_0e}{c^2k^2_{\perp,0}}\left\langle J_0\omega_d\delta H^F_0\right\rangle_E\nonumber\\
&=&-i\frac{c}{B_0}\frac{k_Zk_{\theta,0}}{k^2_{\perp,0}}\left(k^2_Z-k^2_{\perp,0}\right) \frac{\omega_0}{V^2_A}\delta\phi_0\left(\delta\phi_Z-\delta\psi_Z\right).\label{eq:nl_TAE_vorticity}
\end{eqnarray}
The last term on the left hand side is the nonlinear CCT, while the terms on the right hand side are RS\&MX   derived  following the same procedure of equation (\ref{eq:RS_and_MX}). Note that equation (\ref{eq:nl_TAE_vorticity}) has the same structure as its counterpart without EP effects \cite{LChenPRL2012,ZQiuEPL2013}, with the additional physics of the nonlinear CCT as is shown in equation (\ref{eq:nl_EP_TAE}). Ignoring all the nonlinear terms, equation (\ref{eq:nl_TAE_vorticity}) then describes linear TAE excitation by well circulating EP transit resonances \cite{GFuPoFB1989}.

The other equation of TAE, describing the breaking of ideal MHD condition by nonlinear effects, is derived from the parallel component of the nonlinear Ohm's law
\begin{eqnarray}
\delta\phi_0-\delta\psi_0=\frac{c}{B_0}\frac{k_{\theta}}{\omega_0}\delta\phi_0\partial_r\left(\delta\psi_Z-\delta\phi_Z\right).\label{eq:nl_TAE_Ohm}
\end{eqnarray}
Substituting equation (\ref{eq:nl_TAE_Ohm}) into (\ref{eq:nl_TAE_vorticity}), one then reproduces equation (7) of Ref. \cite{LChenPRL2012} in the proper limit, i.e., ignoring EP contributions including wave-particle resonances \cite{ZQiuNF2016} and separating TAEs into finite amplitude pump and its lower/upper sidebands due to radial envelope modulation by ZFZS.
We note that, in the linear growth stage of TAE,   TAE nonlinearity is dominated by the scattering of EP orbit by ZFZS and modification of EP equilibrium (PSZS). While in the  TAE saturation stage, RS\&MX dominates. To understand the more general situations with all the nonlinearities acting on the same footing,  nonlinear equations must be investigated numerically.

\subsection{Nonlinear   EP distribution function evolution}
\label{subsec:PSZS}

It is mentioned in Sec. \ref{subsec:ZFZS} that, $\overline {\delta H^{NL}_Z}$ dominates the PSZS. Noting that $\bar{F}_0=F_0+\overline {\delta H^{NL}_Z}$, we then have, from equation (\ref{NLdc}),
\begin{eqnarray}
\partial_t \bar{F}_0=-i(c/B_0)k_{\theta}J^2_0(k_Z\hat{\rho}_d)\partial_r\overline{\left[J_0\delta L_0\delta H_{0^*}-J_{0^*}\delta L_{0^*}\delta H_0\right]},\nonumber
\end{eqnarray}
and $\bar{F}_0$ can be solved for explicitly using Laplace transformation. Taking
\begin{eqnarray}
\hat{F}_0(\hat{\omega})\equiv \frac{1}{2\pi}\int^{\infty}_0e^{i\hat{\omega}t}\bar{F}_0(t) dt \nonumber
\end{eqnarray}
with $\hat{\omega}$ being the variable for the slow temporal evolution of $\hat{F}_0$, we then have
\begin{eqnarray}
\hat{F}_0(\hat{\omega})&=&\frac{i}{2\pi\hat{\omega}}\bar{F}_0(0)+\frac{c}{B_0}J^2_0(k_Z\hat{\rho}_d)k_{\theta}\frac{1}{\hat{\omega}} \frac{\partial}{\partial r} \int\overline{\left[J_0\delta L_0(y)\delta H_{0^*}(\hat{\omega}-y)-J_{0^*}\delta H_{0^*}(y)\delta H_0(\hat{\omega}-y)\right]}dy.\label{eq:F0_1}
\end{eqnarray}
Here, $\bar{F}_0(0)$ is the initial value of $\bar{F}_0$ at $t=0$. The effects of collisions and external source  can be included in equation (\ref{eq:F0_1}) straightforwardly \cite{FZoncaNJP2015}.  EP response to TAE can also be derived, and we obtain
\begin{eqnarray}
\delta H_{d,0}=-\frac{e}{m}\sum_l J_l(k_{\perp}\hat{\rho}_d) e^{il(\theta-\theta_0)}\int \frac{Q_{0,y'} \hat{F}_0(\hat{\omega}-y')\delta L_0(y')}{\omega_0-k_{\parallel}v_{\parallel}-l\omega_{tr}-ick_Zk_{\theta}\delta L_Z/B_0} dy'.\label{eq:EP_response_laplace}
\end{eqnarray}
Substituting equation (\ref{eq:EP_response_laplace}) into (\ref{eq:F0_1}), we then obtain
\begin{eqnarray}
\hat{F}_0(\hat{\omega})&=&\frac{i}{2\pi\hat{\omega}}\bar{F}_0(0)-\frac{c}{B_0}\frac{e}{m}J^2_0(k_Z\hat{\rho}_d)k_{\theta}\frac{1}{\hat{\omega}} \frac{\partial}{\partial r}\nonumber\\
&\times&\iint \left[\frac{\delta L_0(y)\hat{F}_0(\hat{\omega}-y'-y)\delta L_{0^*}(y')}{y-\omega_0-k_{\parallel}v_{\parallel}-l\omega_{tr}-ick_Zk_{\theta}\delta L_Z/B_0}-
\frac{\delta L_{0^*}(y)\hat{F}_0(\hat{\omega}-y'-y)\delta L_{0}(y')}{\omega_0-y-k_{\parallel}v_{\parallel}-l\omega_{tr}-ick_Zk_{\theta}\delta L_Z/B_0}\right]dydy'.
\end{eqnarray}
Note that  this equation contains $\hat{F}_0$ on both sides as well as wave-particle decorrelation due to ZS in the denominator, and  describes the self-consistent evolution of   EP equilibrium distribution (transport) due to emission and reabsorption of symmetry breaking TAEs. Thus, it corresponds to the Dyson equation in quantum field theory. Its solution  provides the renormalized expression of $\hat{F}_0$ and thus, $\bar{F}_0$. The  nonlinear TAE equation can then be derived, including the  self-consistent interplay between TAE and the EP source, following the derivation for EPM \cite{FZoncaNJP2015}.

For nearly periodic fluctuations, with $\omega_0(\tau)=\omega_{0r}+i\gamma_0(\tau)$, we can assume $\delta \phi_0(t)\equiv\lim_{\tau\rightarrow t}\delta\phi(\tau)\exp(-i\omega_0(\tau)t)$, with $\delta\phi(\tau)=\delta\phi_0\exp(-i\int^{\tau}_0\omega_0(t')dt'+i\omega_0(\tau)\tau)$. Thus, one can show that the Laplace transform
\begin{eqnarray}
\delta\phi(\omega)=\frac{i}{2\pi}\frac{\delta\phi(r,\tau)}{\omega-\omega_0(\tau)}.\nonumber
\end{eqnarray}

Assuming coherent modes with $\gamma_L\ll\omega_{0r}$, we obtain, after some tedious but straightforward algebra
\begin{eqnarray}
\hat{F}_0(\hat{\omega})&=&\frac{i}{2\pi\hat{\omega}}\bar{F}_0(0)-2\frac{c}{B_0}\frac{e}{m}\frac{k_{\theta}}{\hat{\omega}}J^2_0(k_Z\hat{\rho}_d)\sum_l J^2_l(k_{\perp}\hat{\rho}_d)\nonumber\\
&\times&\frac{\partial}{\partial r}\left[\frac{(\hat{\omega}-i\gamma_L)Q_0\hat{F}_0(\hat{\omega}-2i\gamma_L)}{(\omega_{0r}-k_{\parallel}v_{\parallel}-l\omega_{tr}+ick_Zk_{\theta}\delta L_Z/B_0)^2-(\hat{\omega}-i\gamma_L)^2}\left(1-\frac{k_{\parallel}v_{\parallel}}{\omega_{0r}}\right)^2|\delta\phi_0|^2\right].\label{eq:PSZS}
\end{eqnarray}
This expression  contains the information of EP radial transport   and   energy variation due to TAEs, and in the adiabatic limit, reproduces wave-particle trapping \cite{FZoncaNJP2015}. In the simple limits of EPM driven by deeply trapped particle precession resonance, the phase locking between  the radially transported EPs and   frequency sweeping   mode  leads to convective transport of EPs \cite{FZoncaNJP2015}, as referred to as mode particle pumping \cite{RWhitePoF1983}.   A similar picture  is also proposed for the nonlinear saturation of EP-induced geodesic acoustic mode \cite{ZQiuEPS2014}, where phase locking between the pitch angle scattered EPs and the  downward frequency  chirping EGAMs   eventually leads to EP loss due to scattering into lost orbits.

Equations   (\ref{eq:ZFVorticity}), (\ref{eq:zonalcurrent}), (\ref{eq:nl_TAE_vorticity}), (\ref{eq:nl_TAE_Ohm}) and (\ref{eq:PSZS}), thus,  provide the set of nonlinear equations describing the nonlinear evolution of a single toroidal mode number TAE, including both wave-wave nonlinearities and wave-particle phase space nonlinearities. For an in-depth understanding of the nonlinear process, active interactions between analytical theory and large scale  simulations based on first principles are needed.  As a simple application, in Sec. \ref{sec:ZFZS_generation}, we will neglect the wave-particle nonlinearities, and focus on the nonlinear ZFZS generation. Several properties of ZFZS generation are discussed, i.e., fine- vs. meso- scale radial structure, spontaneous excitation vs. forced driven and electrostatic (e.s.) ZF vs. electromagnetic (e.m.) ZC. This allows us to illuminate the underlying physics processes and to clarify  the discrepancies between the analytical theory \cite{LChenPRL2012} and numerical simulations \cite{YTodoNF2010,ZWang2016}.

\section{Nonlinear ZFZS generation}
\label{sec:ZFZS_generation}

In this section, a simplified model neglecting $F_{0,E}$  temporal evolution is considered to investigate the different roles played by EPs and thermal particles  in ZFZS generation.   The spontaneous excitation and forced driven processes are discussed, respectively, in Refs. \cite{LChenPRL2012} and \cite{ZQiuPoP2016}. In this paper,  equations  (\ref{eq:ZFVorticity}) to (\ref{eq:nl_TAE_Ohm}) are derived as the governing equations for ZFZS generation, including both processes presented in Refs. \cite{LChenPRL2012,ZQiuPoP2016}, with the $\delta H^{F}_0$ in equation (\ref{eq:nl_TAE_vorticity}) replaced by its linearized expression. Here, we will show the two limiting cases discussed respectively in Refs. \cite{LChenPRL2012,ZQiuPoP2016}; that is forced driven excitation of ZFZS \cite{ZQiuPoP2016}, which is expected to dominate the early phase of linear instability, and the spontaneous emission of ZFZS \cite{LChenPRL2012}, which is expected to take over the nonlinear  dynamics at later times, after TAE fluctuation amplitude has exceeded a critical  threshold value. Wave particle phase space nonlinearities, meanwhile, are expected to be most important in between these two phases.

\subsection{ZFZS forced driven by TAE in the linear growth stage}

In the linear growth stage of TAE, with the CCT due to EP response dominating over RS\&MX by thermal particles, one can neglect RS\&MX in equation (\ref{eq:ZFVorticity}),
\begin{eqnarray}
\partial_t\hat{\chi}_{iZ}\delta\phi_Z=i\frac{\pi}{4}\frac{k^2_{\perp}}{k_Z}\hat{K}\hat{G}|\hat{A}_0|^2\sum_m|\Phi_0|^2e^{2\gamma_Lt}.\label{eq:ZFtemporal}
\end{eqnarray}
Here,  $\omega_Z\equiv i\partial_t|_Z$ accounts for temporal evolution, and    $\hat{K}\equiv cn_{0E}T_ik_{\theta}/(B_0n_0m_i\rho^2_i\omega^2_0)$.  Noting that $\partial_t=2\gamma_L$,  we then have
\begin{eqnarray}
\delta\phi_Z=i\frac{\pi}{8}\frac{k^2_{\perp}}{k_Z}\frac{\hat{K}\hat{G}}{\gamma_L\hat{\chi}_{iZ}}|\hat{A}_0|^2\sum_m|\Phi_0|^2e^{2\gamma_Lt}.\label{eq:ZF}
\end{eqnarray}
Taking $\Phi_Z\equiv|\Phi_0|^2$ as the ZF fine-scale structure in equation (\ref{def:phiz}), the meso-scale radial envelope of ZF is then
\begin{eqnarray}
\hat{A}_Z=i\frac{\pi}{8}\frac{k^2_{\perp}}{k_Z}\frac{\hat{K}\hat{G}}{\gamma_L\hat{\chi}_{iZ}}|\hat{A}_0|^2.
\end{eqnarray}
This is a typical forced driven process, with the growth rate of the zonal scalar potential being twice that of TAE, and its amplitude proportional to the TAE intensity.
ZC can also be forced driven by TAE. It can be readily estimated from equation (22) of Ref. \cite{LChenPRL2012} that  the amplitude of $\delta\psi_Z$ is much smaller than $\delta\phi_Z$, in that, compared to the case considered in Ref. \cite{LChenPRL2012}, the ZF term is enhanced due to EP response while the ZC term is weakened (frequency mismatch $|\Delta_T|$ replaced by $|2\gamma_L|$ in the ZC term).

Note that, for the forced driven case,  $|\omega_Z|=2\gamma_L$ and $|k_Z|=2|\partial_r\ln\delta\phi_0|$ are fully determined by the linear spectrum of TAE, such that TAE nonlinear equations are not needed to close the system \cite{ZQiuPoP2016}. While for the spontaneous excitation case \cite{LChenPRL2012}, both TAE sidebands and ZFZS equations are needed for taking into account the reinforcement by nonlinearity of the envelope modulation \cite{LChenPoP2000}.

\subsection{ZFZS spontaneous excitation by TAE via modulational instability}
\label{sec:MI_fine_scale}

When TAEs are saturated due to wave-particle phase space nonlinearities, the CCT due to EP nonlinearities can be neglected (note that $\hat{G}\propto \gamma_L$), and RS\&MX play the dominant role
in the vorticity equation. Equations (\ref{eq:ZFVorticity})-(\ref{eq:nl_TAE_Ohm}) can then recover the coupled nonlinear ZFZS\&TAE equations derived in Ref. \cite{LChenPRL2012}, leading to ZFZS spontaneous excitation when the conditions for modulational instability are satisfied. The finite coupling comes from radial envelope  induced symmetry breaking \cite{AHasegawaPoF1978,LChenPoP2000}, and thus it is natural to separate the TAEs into a constant amplitude pump and its sidebands due to ZFZS radial envelope modulation \cite{LChenPoP2000}.
The  threshold condition for the modulational instability is determined by the  frequency mismatch of TAE sidebands  ($\Delta_T/\omega_0\sim O(\epsilon)$)  associated with finite envelope modulation.  Separating TAEs into fixed amplitude pump and sideband, i.e., $\delta\phi_0=\delta\phi_P+\delta \phi_+$, $\delta\phi_{0^*}=\delta\phi_{P^*}+\delta \phi_-$,
\begin{eqnarray}
\delta\phi_P&=&A_0e^{i(n\phi-m_0\theta-\omega_0t)}\sum_j e^{-ij\theta}\Phi_0(x-j),\nonumber\\
\delta\phi_{\pm}&=&A_{\pm}e^{\pm i(n\phi-m_0\theta-\omega_0t)}e^{i(\int \hat{k}_Zdr-\omega_Zt)}\nonumber\\
&&\hspace*{2em}\times\sum_je^{\mp ij\theta}\left\{\Phi_+(x-j) \atop \Phi_-(x-j)\right\}.\nonumber
\end{eqnarray}
Here, $\theta_{k,P}=0$ is assumed for the pump TAE. Since the spontaneous excitation process dominates when EP resonant drive is very weak,  $\partial_r\ln\Phi_{0}-c.c.$ in $\hat{F}$ vanishes, and the only symmetry breaking mechanism (to have $\hat{F}\neq0$) is finite envelope modulation.  As a result,
\begin{eqnarray}
\frac{\partial^2}{\partial r^2}\hat{F}|A_0|^2\sum_m|\Phi_0|^2&=&\frac{\partial^2}{\partial r^2}\hat{F}\overline{\left(\delta\phi_P\delta\phi_-+\delta\phi_{P^*}\delta\phi_+\right)}\nonumber\\
&=&i\hat{k}_Zk^2_Z\overline{\left(\delta\phi_P\delta\phi_--\delta\phi_{P^*}\delta\phi_+\right)}\nonumber
\end{eqnarray}
and, thus, equation (\ref{eq:RS_and_MX}) becomes
\begin{eqnarray}
\mbox{RS+ MX}&=&-i\frac{c}{B_0}\frac{n_0e^2}{T_i}k_{\theta}\hat{k}_Zk^2_Z\rho^2_i\frac{1}{\omega_Z}\left(1-\frac{k^2_{\parallel,0}v^2_A}{\omega^2_0}\right) \left(A_PA_--A_{P^*}A_+\right)\sum_m|\Phi_0|^2. \label{eq:RS_and_MX_old}
\end{eqnarray}
Noting that $\omega_Z=i\partial_t$ and $k^2_Z=\left(\hat{k}_Z-i\partial_r\sum_m\ln|\Phi_0|^2\right)^2$ from the balance of the radial variations on both sides of the ZFZS vorticity equation, equation (\ref{eq:RS_and_MX_old}) can then recover equation (3) of Ref. \cite{LChenPRL2012} after averaging over parallel mode structures ($\sum_m |\Phi_0|^2=1$ normalization is assumed in Ref. \cite{LChenPRL2012}, which applies to high/moderate ballooning drift waves \cite{FZoncaPoP2004}, while it may not be generally valid for TAEs). Taking into account the fine-scale structure of ZFZS, and keeping only the dominant poloidal harmonic of TAE, we then obtain
\begin{eqnarray}
i\omega_Z\hat{\chi}_{iZ}\delta\phi_Z= -\frac{c}{B_0}k_{\theta}\hat{k}_Z\left(1-\frac{k^2_{\parallel}V^2_A}{\omega^2_0}\right)\left(A_PA_--A_{P^*}A_+\right)|\Phi_0|^2.\label{eq:ZF_1}
\end{eqnarray}

The zonal current equation can be derived from  equation (\ref{eq:zonalcurrent}), noting that $\partial_r|\delta\phi_0|^2=\partial_r\delta\psi_Z$,
\begin{eqnarray}
\delta\psi_Z=-i\frac{c}{B_0}\frac{k_{\theta}}{\omega_0}k_Z\left(A_PA_-+A_{P^*}A_+\right)|\Phi_0|^2.\label{eq:ZC_1}
\end{eqnarray}

Neglecting the contribution of EPs, we then obtain, from equations (\ref{eq:nl_TAE_vorticity}) and (\ref{eq:nl_TAE_Ohm}), the nonlinear equations for TAE upper/lower sidebands,
\begin{eqnarray}
\left(k^2_{\parallel}V^2_A\delta\psi-\omega^2\delta\phi+\omega^2_G\delta\phi\right)_{\pm}&=&i\frac{c}{B_0}\frac{k_Zk_{\theta,0}}{k^2_{\perp,\pm}}\left(k^2_Z-k^2_{\perp,0}\right) \omega_0\left\{\delta\phi_P \atop \delta\phi_{P^*}\right\}\left(\delta\phi_Z-\delta\psi_Z\right),\label{eq:TAE_sideband_vorticity} \\
(\delta\phi-\delta\psi)_{\pm}&=&i\frac{c}{B_0}\frac{k_{\theta}k_Z}{\omega_0}\left\{\delta\phi_P \atop \delta\phi_{P^*}\right\}\left(\delta\psi_Z-\delta\phi_Z\right).\label{eq:TAE_sideband_ohm}
\end{eqnarray}
Substituting equation (\ref{eq:TAE_sideband_ohm}) into (\ref{eq:TAE_sideband_vorticity}), noting that $k^2_{\parallel}V^2_A\simeq \omega^2$, $|k_r|\simeq|\partial_r\ln\Phi_0|\gg|k_{\theta}|$ for pump TAE in the inertial layer,   $|k_Z|\simeq |\partial_r\ln\Phi_Z|=2|\partial_r\ln\Phi_0|$ and $|k^2_{\perp,\pm}|\simeq(|\partial_r\Phi_0|+|\partial_r\Phi_Z|)^2\simeq 9|\partial_r\Phi_0|^2$, we then have
\begin{eqnarray}
\left(k^2_{\parallel}V^2_A- \omega^2 + \omega^2_G\right)k^2_{\perp,\pm}\delta\phi_{\pm}= 6i\frac{c}{B_0} k_{\theta}k_Zk^2_{r,0} \omega_0\left\{\delta\phi_P \atop \delta\phi_{P^*}\right\}(\delta\phi_Z-\delta\psi_Z).\label{eq:TAE_sideband_1}
\end{eqnarray}
Note that, in the present work, $(r,\theta,\phi)$ is assumed as a right-handed coordinate, and the nonlinear terms in equations (\ref{eq:ZF_1}), (\ref{eq:ZC_1}) and (\ref{eq:TAE_sideband_1}) have opposite sign to Ref. \cite{LChenPRL2012}. Assuming $\Phi_0\equiv \exp(-x^2/(2\Delta^2_r))/(\pi^{1/4}\Delta^{1/2}_r)$ with $\Delta_r\sim O(\sqrt{\epsilon})$ being the characteristic scale length of the fine structure,   $\Phi_Z=|\Phi_0|^2$,   defining $\mathscr{E}_T\equiv\langle\langle L_T\rangle\rangle$, with $L_T\equiv k^2_{\parallel}V^2_A+\omega^2_G -\omega^2_T$,   $\langle\langle\cdots\rangle\rangle\equiv\int (\cdots)|\Phi_0|^2dx$, and noting $\Phi_{\pm}=\Phi_0$ to the leading order, we then have
\begin{eqnarray}
k^2_{\pm,\perp}\mathscr{E}_{\pm}A_{\pm}&=&6i\frac{c}{B_0}k_{\theta}\omega_0\left\{A_P \atop A_{P^*}\right\}(A_Z-\Psi_Z)\langle\langle k_Z|\partial_r\Phi_0|^2\rangle\rangle\label{eq:TAE_sideband_equation},\\
\Psi_Z&=&-i\frac{c}{B_0}\frac{k_{\theta}\hat{k}_Z}{\omega_0}(A_PA_-+A_{P^*}A_+)\label{eq:zonal_current_envelope},\\
i\omega_Z\hat{\chi}_{iZ}A_Z&=&-\frac{c}{B_0}k_{\theta}\hat{k}_Z\langle\langle 1- k^2_{\parallel}V^2_A/\omega^2_0\rangle\rangle(A_PA_--A_{P^*}A_+)\label{eq:zonal_flow_envelope}.
\end{eqnarray}
Here, $\Psi_Z$ is the radial envelope of $\delta\psi_Z$, $\mathscr{E}_{\pm}=\left(\omega^4_A\Lambda_T(\omega)D(\omega,k_Z)/(\epsilon_0\omega^2)\right)_{\omega=\omega_{\pm}}$,  $\omega_A\equiv V_A/(qR_0)$, $\Lambda_T=\sqrt{-\Gamma_+\Gamma_-}$ with $\Gamma_{\pm}=(\omega^2/\omega^2_A-1/4)\pm\epsilon_0\omega^2/\omega^2_A$, $D(\omega,k_Z)=\Lambda_T-\delta \hat{W}_k(\omega,k_Z)$, and $\delta \hat{W}_k(\omega,k_Z)$ plays the role of a normalized potential energy \cite{FZoncaPoFB1993,FZoncaPoP2014a,FZoncaPoP2014b}.
Substituting $A_{\pm}$ from equation (\ref{eq:TAE_sideband_equation}) into equations (\ref{eq:zonal_current_envelope}) and (\ref{eq:zonal_flow_envelope}), letting $\omega_Z= i\gamma_Z$ and noting
$D_{\pm}=\pm(\partial D_0/\partial\omega_0)(i\gamma_Z\pm\Delta_T)$ with $\Delta_T=\omega_T(k_r)-\omega_0$ being the frequency mismatch, we then obtain
\begin{eqnarray}
A_Z&=&-6\left(\frac{c}{B_0}\frac{k_{\theta}}{k_{\perp}}\right)^2\frac{\omega_0}{\hat{\chi}_{iZ}}\hat{k}_Z\langle\langle k_Z|\partial_r\Phi_0|^2\rangle\rangle \frac{\langle\langle 1- k^2_{\parallel}V^2_A/\omega^2_0\rangle\rangle \epsilon_0\omega^2_A }{\omega^4_A\Lambda_T(\omega)\partial D_0/\partial\omega_0 (\gamma^2_Z+\Delta^2_T)}|A_0|^2\left(A_Z-\Psi_Z\right)\nonumber\\
&\equiv&-\hat{\alpha}_{\Phi}(A_Z-\Psi_Z)/(\gamma^2_Z+\Delta^2_T)\label{eq:zonal_flow_envelope1},\\
\Psi_Z&=&6\left(\frac{c}{B_0}\frac{k_{\theta}}{k_{\perp}}\right)^2\hat{k}_Z\langle\langle k_Z|\partial_r\Phi_0|^2\rangle\rangle |A_0|^2\frac{2\Delta_T\epsilon_0\omega^2_A}{\omega^4_A\Lambda_T(\omega)\partial D_0/\partial\omega_0 (\gamma^2_Z+\Delta^2_T)}\left(A_Z-\Psi_Z\right)\nonumber\\
&\equiv&-\hat{\alpha}_{\Psi}(A_Z-\Psi_Z)/(\gamma^2_Z+\Delta^2_T)\label{eq:zonal_current_envelope1}.
\end{eqnarray}
Note that, equations (\ref{eq:zonal_flow_envelope1}) and (\ref{eq:zonal_current_envelope1})  correspond, respectively,  to equations (19) and (20) of Ref. \cite{LChenPRL2012}; and the coefficients $\hat{\alpha}_{\Phi}$ and $\hat{\alpha}_{\Psi}$ correspond to $\alpha_{{\Phi}T}$ and $\hat{\alpha}_{{\Psi}T}$ of Ref. \cite{LChenPRL2012}, with the enhanced coupling due to inclusion of ZFZS fine radial structure taken into account \cite{ZQiuNF2016}. The equations in Ref. \cite{LChenPRL2012} can be recovered by replacing $\langle\langle k_Z|\partial_r\Phi_0|^2\rangle\rangle$ with $\hat{k}_Zk^2_{\theta}/3$ and $k_{\perp}$ with $(\hat{k}^2_Z+k^2_{\theta})^{1/2}$. The nonlinear dispersion relation of the modulational instability can then be derived as:
\begin{eqnarray}
\gamma^2_Z=\hat{\alpha}_{\Psi}-\hat{\alpha}_{\Phi}-\Delta^2_T.
\end{eqnarray}

The condition for the modulational instability is given by
\begin{eqnarray}
\hat{\alpha}_{\Psi}-\hat{\alpha}_{\Phi}>\Delta^2_T.
\end{eqnarray}
Thus, the threshold condition on pump TAE amplitude is much lower due to the enhanced nonlinear coupling, while the conditions for e.s. ZF or e.m. ZC to be preferentially excited is exactly the same as that discussed in Ref. \cite{LChenPRL2012}.  Assuming the condition for ZC excitation is satisfied ($\Delta_T/\omega_0>0$) \cite{LChenPRL2012},  the threshold on pump TAE amplitude for ZC spontaneous excitation is lower by $\sqrt{k^2_{\theta}/(3\langle\langle  |\partial_r\Phi_0|^2\rangle\rangle)}\sim O(\epsilon)$ due to the inclusion of ZFZS fine scale structures.

\section{Discussions and Summary}
\label{sec:summary}

The different properties of the nonlinear processes, e.g., fine- vs. meso- radial scale, forced driven vs. spontaneous excitation, can be illuminated from our derivations and theoretical analysis in Sec. \ref{sec:ZFZS_generation}.  Below, we discuss them one by one.

\subsubsection{Forced driven vs. spontaneous excitation}

In the linear growth stage of the pump TAE, there is an $e^{2\gamma_L t}$ factor in the nonlinear terms due to the coupling of the pump TAE to its complex conjugate. The operator for temporal evolution $\omega_Z$ is then $\omega_Z=2i\gamma_L$, while the driven ZF amplitude is proportional to the intensity of pump TAE. This is a typical forced driven process.
On the other hand, as TAE saturates due to wave-particle nonlinearities, the CCT contribution becomes negligible, and finite RS\&MX requires finite radial envelope modulation \cite{AHasegawaPoF1978} (i.e., the sidebands assumed in Ref. \cite{LChenPRL2012}).  As a result,   nonlinear equations for sidebands are needed to close the system; and for analyzing spontaneous excitation of ZFZS.

This clarifies the discrepancy of simulations \cite{YTodoNF2010,ZWang2016} and analytical theory based on modulational instability of a pump TAE with a prescribed amplitude \cite{LChenPRL2012}. In the simulations, the TAEs are driven by EPs, and the observed forced driven process occurs in the linear growth stage of TAE \cite{YTodoNF2010}. To observe the spontaneous excitation process, one has to wait long enough till RS\&MX are comparable to/larger than the CCT; and one has to be careful in distinguishing the underlying nature  of the different zonal components.
One possible way to clearly demonstrate the spontaneous excitation process, is to get   a stationary pump TAE with constant amplitude by antenna or by carefully posing an artificial dissipation to balance the EP drive.

\subsubsection{Fine-  vs. meso-scale structures}

The radial structure of the generated ZF component of the ZS is given in equation (\ref{eq:ZFVorticity}), and the radial variation can be from either the meso-scale radial envelope ($|A_0|^2$) or the fine-scale parallel mode structure ($\sum_m |\Phi_0|^2$) of pump TAE. Note that equation describing ZF excitation by drift waves (DWs) has a similar structure \cite{LChenPoP2000}, but ZF excited by DWs typically has a meso-scale structure. In fact, for DWs characterized with moderate or strong ballooning structure, $\sum_m |\Phi_0|^2=1$ \cite{LChenPoP2000,FZoncaPoP2004}, and   radial variation comes from $|A_0|^2$.   On the other hand, AEs are typically weakly ballooning due to the presence of SAW continuum, such that $\sum_m |\Phi_0|^2$ dominates radial variation. As a result, ZF driven by TAE (more generally, AEs) has a  fine-scale radial structure, in addition to the well-known meso-scale envelope. The same argument and considerations apply for the ZC component of the ZS, expressed by equation (\ref{eq:zonalcurrent}).

\subsubsection{Zonal flow vs. zonal current}

The condition for ZF or ZC spontaneous excitation has been discussed in detail in Ref. \cite{LChenPRL2012}. ZF and ZC generation are described by, respectively, vorticity equation and Ohm's law.
It is shown in equation (22) of Ref. \cite{LChenPRL2012} that, for pump TAE with given amplitude, ZF generation is screened by neoclassical shielding and limited RS\&MX near-cancelation; while ZC generation is related to frequency mismatch. For certain plasma equilibria, ZC generation has a much lower threshold condition.  On the other hand, with EP effect taken into account, ZF generation can dominate since CCT due to EP contribution is much larger than RS\&MX, as we shown in Sec. \ref{subsec:ZFZS}. This explains why e.s. ZF generation is always observed in the simulations \cite{YTodoNF2010,ZWang2016,HZhangPST2013}. Furthermore, note that in Ref. \cite{YTodoNF2010}, the bulk plasma is treated by MHD model such that neoclassical shielding is not accounted for, and ZF is further enhanced. To observe ZC, one has to run the simulation longer till EP effects are weakened by, e.g., wave-particle trapping. Plasma equilibrium must also be compatible with ZC excitation conditions \cite{LChenPRL2012}.

In conclusion, the set of equations describing  nonlinear evolution of a single toroidal mode number TAE are derived, including both $n=0$ ZFZS generation and $n=0$ wave-particle phase space nonlinearities. A simplified case neglecting wave-particle phase nonlinearity is then investigated to study the different roles of EPs and bulk plasma  on ZFZS generation. The EP and bulk plasma contribution are derived on the same footing, and we show that, due to their different orbit sizes, EP contribution dominates in the ideal region of TAE while bulk plasmas dominates in the inertial layer. On the other hand, due to their different mechanisms to break the Alfv\'enic state \cite{LChenPoP2013}, EP contribution dominates in the linear growth stage of the pump TAE, while bulk plasma contribution takes over as the pump TAE   saturates by wave-particle phase space nonlinearity. Consequently, the different properties of ZFZS generation observed in numerical simulations, e.g., forced driven vs. spontaneous excitation, fine- vs. meso- scale radial structure and e.m. ZC vs. e.s. ZF, can be understood and explained  within the present theoretical analysis.

\section*{Acknowledgments}
This work is supported by     US DoE GRANT,  the National Magnet Confinement Fusion Research Program under Grants Nos.  2013GB104004  and   2013GB111004, the National Science Foundation of China under grant Nos.  11575157  and 11235009,  Fundamental Research Fund for Chinese Central Universities under Grant No. 2016FZA3003 and   EUROfusion Consortium
under grant agreement No. 633053.

\end{document}